\documentclass[preprintnumbers,showkeys,floats,prd,aps,nofootinbib]{revtex4}
\usepackage{tipa}
\usepackage{amssymb,amsmath}
\usepackage[dvips]{graphicx}
\usepackage{graphicx}
\usepackage{amsmath}
\usepackage{amssymb}
\usepackage{epsfig}
\usepackage{indentfirst}
\usepackage{float}
\usepackage{multirow}
\usepackage[usenames]{color}

\def\be{\begin{equation}}
\def\ee{\end{equation}}
\def\ba{\begin{eqnarray}}
\def\ea{\end{eqnarray}}

\def\@eqnnum{{\normalfont \normalcolor [\theequation]}}
\makeatother \makeatletter
\begin{document}

\title{Scale Invariance from Modified Dispersion Relations}
\author{Yao Lu}
\email{aliseturtle@gmail.com}
\author{Yun-Song Piao}
\email{yspiao@gucas.ac.cn} \affiliation{College of Physical
Sciences, Graduate School of Chinese Academy of Sciences, Beijing
100049, China }

\begin{abstract}
In this paper, inspired by the investigations on the theory of
cosmological perturbations in Ho\v{r}ava-Liftshit cosmology, we
calculated  the spectrum of primordial perturbation leaded by a
scalar field with modified dispersion relation $\omega\sim
k^z/a^{p-1}$, in which $z$ is the critical exponent and $p$ is
generally not equal to $z$. We discussed that for fixed $z$, if the
spectrum is required to be scale invariant,
how should $p$ depend on the background evolution. We concluded
that there is always a room of parameters for the generation of
scale invariant spectrum.



\end{abstract}

\preprint{} \maketitle

\baselineskip=16pt

\section{Introduction}

Inflation, e.g. see \cite{inf}, is the most prevailing scenario for
early universe since it first provides an solution for the horizon,
flatness and entropy problems of SBB. While the most important
success of inflation is it suggests a causal generation mechanism
for the superhorizon, adiabatic, scale invariant density
perturbation.
However, the causal generation of primordial perturbation can also
be implemented in the case with the decaying speed of sound
\cite{arm4},\cite{armm},\cite{piao07},\cite{magueijo}, and also
\cite{khoury} for contracting phase. During an earlier period of
universe, when the sound speed was decaying, the sound horizon
contracted, thus the causal perturbation initially deep into the
sound horizon can emerge and be stretched to the super sound horizon
scale, and become the primordial perturbation responsible for
structure formation at late time. In such a case, in principle,
inflation may not be required at least for the generation of
primordial perturbation. The modification of sound speed actually
corresponds to the modification of dispersion relation. The
cosmological applications of modified dispersion relation have been
studied, which include
trans-Planckian physics \cite{transp}, noncommutative field approach \cite%
{noncom}, loop quantum gravity \cite{loop}, Lorentz-violating models \cite%
{lv}.

Recently, a candidate for quantum gravity, dubbed
Ho\v{r}ava-Liftshitz (HL) gravity \cite{H}, has been studied
intensively. This theory has an action that is power counting
renormalizable with respect to a scaling symmetry which treats space
and time differently. In the IR region, this theory naturally flows
to general
relativity. The black hole solutions in Ho\v{r}ava gravity were studied in \cite%
{horavabh},\cite{CCO},\cite{MK},\cite{KS},\cite{CCO1},\cite{CJ},\cite{LKM}.
The cosmological solutions was first addressed in
\cite{horavacos},\cite{KK},\cite{LMP}. It was found that the early
universe in HL cosmology may be able to escape singularity and has
a nonsingular bounce. This might give an alternative to inflation,
as has been discussed in \cite{matbounce},\cite{CS}. There has
been many detailed analysis for the theory of cosmological
perturbations in HL cosmology
\cite{tokyoguy},\cite{piao09},\cite{Gao},\cite{CHZ},\cite{CPT},\cite{gaoxian},\cite{CZ},\cite{YKN},\cite{BS},\cite{WM},
and also \cite{TS},\cite{Koh} for gravitational wave. In
\cite{tokyoguy}, it was argued in UV regime of HL gravity which
has a critical exponent $z=3$, the spectrum of primordial
perturbation induced by a scalar field may be scale invariant for
any power law expansion with $a\thicksim t^{n}$ and $n>1/3$. This
was subsequently confirmed by solving the motion equation of
perturbation mode on super sound horizon scale for any background
evolution of early universe \cite{piao09}. In \cite{piao09}, not
only how the primordial perturbation generated in UV regime is
matched to the observations on large IR scale was illustrated, but
also the dependence of spectral index on any $z$ and $n$ was
showed, in addition, it was also pointed out that the case of
$n<1/3$ for the generation of scale invariant spectrum actually
corresponds to that in contracting phase. There are also many
studies on other aspects of HL cosmology
\cite{MNTY},\cite{CJG},\cite{TN},\cite{S1},\cite{M1},\cite{M2},\cite{MIP},
and other relevant issues
\cite{V},\cite{K1},\cite{CH},\cite{SVW},\cite{GC},\cite{M3}, and
some challenges \cite{CNPS},\cite{LP}.


In this paper, inspired by the investigations on the theory of
cosmological perturbations in HL cosmology, we will study the
spectrum of primordial perturbation induced by a scalar field with
modified dispersion relation $\omega\sim {k^z\over a^{p-1}}$, in
which $z$ is the critical exponent. a specific case of our work with
$p=z$ have been studied in \cite{CZ}. We will generalize the
calculations of \cite{piao09} to the case with arbitary $p$ and then
give our discussion on the scale invariance of spectrum. The rest of
the paper is organized as follows. In sec.2 we will calculate the
spectrum of primordial perturbation leaded by a scalar field with
modified dispersion relation $\omega\sim {k^z\over a^{p-1}}$. In
sec.3 we will discuss that for $z$ ranged from 1 to 4, if the
spectrum is required to be scale invariant, what should the value of
$p$ be. We will depict the parameter space that guarantees scale
invariant spectrum. The final is our conclusion.

\section{primordial perturbation from modified dispersion relations}

We will briefly calculate the primordial perturbation induced by a
scalar field with modified dispersion relation $\omega\sim
{k^z\over a^{p-1}}$. In the momentum space, the motion equation of
perturbation of scalar field $\varphi$ is given by \be
u_k^{\prime\prime} +\left(\omega^2-{a^{\prime\prime}\over
a}\right) u_k = 0 ,\label{equ}\ee where $u_k$ is related to the
perturbation $\delta\varphi$ of $\varphi$ by $a\delta\varphi_k =
\frac{1}{\sqrt{2}}[a_{ \overrightarrow{k}}^{-}u_{k}^{\ast }(\eta
)+a_{-\overrightarrow{k} }^{+}u_{k}(\eta )]$, $a_{
\overrightarrow{k}}^{-}$ and $a_{-\overrightarrow{k} }^{+}$ are
mode annihilation and generation operators. and the prime denotes
the derivative with respect to the conformal time $\eta$. We can
define $\Omega ^{2}(\eta )=\omega^2-\frac{a^{\prime \prime }}{a}$
for following discussions. Here, $\omega$ is taken as \be
\omega={k^z\over a^{p-1}M^{z-1}}. \label{omega}\ee When $p=z$, the
case is same as that studied in \cite{piao09}. However, here we
will regard $p$ as arbitrary value, which may be equal to $z$ and
may be not. Recently, in the studies on the theory of cosmological
perturbations in HL cosmology,
e.g.\cite{CHZ},\cite{CPT},\cite{gaoxian}, it has been found that
there are such some terms, which might imply that the factor
before ${k^z\over a^{z-1}M^{z-1}}$ has a dependence on the time,
i.e. $\omega\sim f(\eta){k^z\over a^{z-1}M^{z-1}}$. This actually
can be rewritten as (\ref{omega}), since $a$ is the function of
$\eta$. It is this observation that motivates our study.

The calculations of perturbation spectrum is actually a
generalization of that in \cite{piao09}. In term of (\ref{omega}),
the sound speed is given by $c_s={k^{z-1}\over a^{p-1}M^{z-1}}$ and
the sound horizon is ${c_s\over H}$, where $H$ is the Hubble
parameter. In this case, how the primordial perturbation generated
in UV regime of HL gravity is matched to the observations on large
IR scale has been illustrated in e.g. Fig.1 of \cite{piao09}. We
assume that initially all modes of fluctuation starts in their
vacuum state. But in order to define vacuum in analogy with that in
flat spacetime, adiabatic condition should be admitted. We need
$\Omega $ change quite small during one period $\sim\Omega $ of
oscillation,i.e. $\frac{|\Omega ^{\prime }|}{\Omega ^{2}}<<1$. In
this case, the mode function is approximatelly that in Minkovsky
spacetime,
\begin{equation}
u_{k}\approx \frac{1}{\sqrt{2\Omega (k,\eta )}}\exp\left(
-i\int^{\eta
}\Omega (k,\eta ^{\prime })d\eta ^{\prime }\right) ,~~~\text{when }~\frac{%
\left\vert \Omega \right\vert ^{\prime }}{\Omega ^{2}}\ll 1
\label{adiabatic}
\end{equation}
notice $\Omega\approx\omega$ here, since we need
$\frac{a''}{a}\ll\omega^{2}$(equivalent to adiabatic condition) to
validate the approximation, i.e. the perturbation mode is deep into
sound horizon.

However, what we concern is that with $\omega\eta\ll 1$, i.e. on
super sound horizon scale, which is directly related with our
observation. This can be obtained by solving Eq.(\ref{equ}) with the
initial condition (\ref{adiabatic}), as has been done in the
calculations of inflationary perturbation. We take a power law
evolving background $a(t)=a_{*}({t\over t_*})^{n}$ for study, where
$n$ is assumed to be positive real constant, and $t_{*}$ is regarded
as a reference time and the corresponding scale factor is
$a_{*}$. In conformal time, $a=a_{*}(\frac{\eta }{\eta _{*}})^{(\frac{n%
}{1-n})}$ where $\eta_{*}=t_{*}(1-n)$. Eq.(\ref{equ}) can be reduced
to Bessel equation. Thus with the initial condition
(\ref{adiabatic}), its solution is
\begin{equation}
u_{k}=0.5\sqrt{\frac{\pi |\eta |(1-n)}{np-1}}{\cal H}
_{v}^{(1)}\left( \left\vert \frac{(n-1)\omega \eta}{np-1}
\right\vert \right),
\end{equation}%
where ${\cal H} _{v}^{(1)}$ is the first kind of Hankel function
with order \be v=0.5\left|{3n-1\over np-1}\right|. \label{v}\ee
First we will check the time dependence of mode amplitude in
different cases. When $\omega\eta\rightarrow 0$ at late time,
\begin{equation}
{\cal H}_{v}^{(1)}\left( \left\vert \frac{(n-1)\omega \eta}{np-1}
\right\vert \right) \approx -i\left( 2\left\vert
\frac{np-1}{(n-1)\omega \eta }\right\vert \right) ^{v}\frac{\Gamma
(v)}{\pi }
\end{equation}
is given. Thus in term of definition, vacuum expectation
$\langle\delta \varphi_k^2\rangle={|u_k|^2\over 2a^2}$, we have
\begin{equation}
\delta\varphi  \propto |\eta |^{\alpha },
\end{equation}
where $\alpha =v(\frac{np-1}{1-n})- 0.5(\frac{3n-1}{1-n})$. Thus
dependent on the sign of ${3n-1\over pn-1}$, $\delta\varphi_k$
evolves in 2 different cases. We can see that $\alpha =0$ when
${3n-1\over pn-1}>0$. In this case, $\delta\varphi_k$ is constant on
super horizon scale, which corresponds to the constant mode. While
${3n-1\over pn-1}<0$, $\alpha =\frac{3n-1}{n-1}$. In this case, the
amplitude will change with time, which can be increasing or
decaying, dependent on different background evolution. For example,
if $n>1$, $\alpha
>0$, and $|\eta |$ increases in contracting phase and decays in
expansion phase, thus $\delta\varphi $ increases in contracting
phase and decrease in expansion phase, since $\eta$ runs initially
from $-\infty$ to $0_-$ in the expanding phase with $n>1$, whereas
in contracting phase $\eta$ runs initially from $0_+$ to $\infty$.
The various cases are listed in the table.

\begin{table}[htbp]
\centering 
\begin{tabular}{|c|c|c|}
\hline & expansion & contraction \\ \hline $n>1$ & decay & increase
\\ \hline $1/3<n<1$ & decay & increase \\ \hline $n<1/3$ &
increase & decay \\ \hline
\end{tabular}%
\end{table}

In next section, it will be showed that $np>1$ in expanding phase
and $np<1$ in contracting phase are required for the emergence of
primordial perturbation, i.e. initially the perturbation mode is
deep into the sound horizon, and then is stretched to the super
sound horizon scale. This condition together with
$\frac{3n-1}{np-1}<0$ or $\frac{3n-1}{np-1}>0$ will prohibit some
possibilities. For example, since $\frac{3n-1}{np-1}<0$, if $
n>1/3$, then $p<1/n$, in this case only the contracting phase is
possible, while the decaying solution in the expanding phase is
prohibited since the perturbation can not emerge in the expanding
phase with $np<1$.

The perturbation spectrum is given by
\begin{equation}
\mathcal{P}_{\varphi }=\beta H_{\ast}^{2}\left( \eta H_{\ast}\right)
^{2\alpha }\left( \frac{M}{H_{\ast}}\right) ^{(2z-2)v}\left(
\frac{k}{H_{\ast}}\right) ^{n_s-1}
\end{equation}%
where the constant $\beta$ is \begin{equation}
\beta=\frac{\left( \Gamma \left( v\right) \right) ^{2}}{4\pi ^{3}}%
\left[ 2\left(\frac{np-1}{1-n}\right)\right] ^{2v-1}\left( \frac{n}{1-n}\right) ^{\frac{%
2n-2npv+2nv}{1-n}}
\end{equation}
and $a_*$ has set as 1 and $H_*$ denotes the Hubble scale at the end
of perturbation generation period. We can see that for $z=1$ and
$n\gg 1$, i.e. the scalar field with normal dispersion relation
during inflation, $\mathcal{P}_{\varphi }\sim H_{\ast}^{2}$ can be
obtained, since $\alpha=0$ and $n_s=1$, which is consistent with
well known result. For $p=z=3$, we have $\alpha=0$, $v=0.5$ and
$n_s=1$, thus $\mathcal{P}_{\varphi }\sim M^{2}$, which is
consistent with that obtained in \cite{tokyoguy},\cite{piao09}. The
spectral index $n_s$ is
\begin{equation}
n_{s}-1=3-z\left\vert \frac{3n-1}{np-1}\right\vert
\label{ns}\end{equation}%
When $p=z=1$, (\ref{ns}) will reproduce the result obtained with
normal dispersion relation, i.e. the spectrum is scale invariant
only when $n\gg 1$(inflation) or $n={2\over 3}$(contraction with
$w\simeq 0$) \cite{Wands99},\cite{FB},\cite{S}. 
While when $p=z$, the result in \cite{piao09} is recover, in which
the parameter space of scale invariance of spectrum is studied in
details. However, here since $p$ can be generally not equal to $z$,
it can be expected that the parameter space for scale invariance
could be enlarge, i.e. there can be more possibilities to obtain the
scale invariant spectrum, which will be investigated fully in the
following section.

\section{Parameter space for scale invariance}

In general, the fluctuations of field can only seed primordial
perturbations inside Hubble radius, otherwise no stable vacuum of
$\delta \varphi_k$ can be defined. The generation of an adiabatic,
Gaussian and scale invariant spectrum of superhubble perturbations
can not happen in any period of standard big bang. It is generally
argued that only during a period of accelerated expansion, in which
$a/k$ grows faster then ${1\over H}$, can subhubble modes being
stretched to become superhubble modes. In this case fluctuations can
be generated when they were subhubble and become superhubble during
inflation.
However, in the general cases with a modified dispersion relation,
the argument should be changed as follows. The adiabatic condition
for Eq.(\ref{equ}) is $\frac{|\Omega ^{\prime }|}{\Omega ^{2}}\ll1$,
which is satisfied when $\omega ^{2}\gg a^{\prime \prime }/a$, since
in this case $\frac{|\Omega
^{\prime }|}{\Omega ^{2}}=\frac{|\omega ^{\prime }|}{\omega ^{2}}$ and $%
\frac{|\omega ^{\prime }|}{\omega ^{2}}\ll1$ is equivalent to
$\omega ^{2}\gg a^{\prime \prime }/a$. Because $a^{\prime \prime
}/a\sim 1/\eta ^{2}$, we conclude that $|\omega \eta |>>1$  is
needed for the existence of adiabatic vacuum. This result can be
viewed in 2 perspectives. The first one is to define an effective
wavelength as $a/\omega $. The initial condition $|\omega \eta |\gg
1$ corresponds the effective wavelength deep inside the Hubble
horizon at the beginning $a/\omega \ll 1/h$. The second one is to
define a sound horizon as ${c_{s}\over |H|}$, in which $c_{s}=\omega
/k$. $|\omega \eta |>>1$ means the physical wavelength $a/k $ is
much smaller than the sound horizon ${c_{s}\over |H|}$, thus a
casual relationship can be established on super horizon scale. The
details of this argument can be found in \cite{piao09}. The next
step is to analysis the subsequent evolution of $|\omega \eta |$
once initial requirement is satisfied. In power law expansion or
contracting $a\sim |t|^{n}$ universe, $t=0$ is defined to be big
bang singularity.
We have $%
|\omega \eta |\sim |t|^{1-np}$, since $\omega \sim a^{1-p}\sim
|t|^{n(1-p)}$ and $|\eta |\sim |t|^{1-n}$. When our universe starts
with big bang and expands with a power law expansion $a\sim t^{n}$,
$np>1$ is required to guarantee $|\omega \eta |\gg 1$. What follows
is $|\omega \eta |$ decreases monopoly to much smaller than $1$, and
this means corresponding mode leaves the sound horizon. Otherwise,
the requirement that physical wavelength should be deep inside the
sound horizon in the early universe can not be assured. But what if
we have another explanation for `early universe'?  Even if $|\omega
\eta |\ll1$ in the vicinity of big bang, there are still possibility
that during a power law contracting phase, our universe starts with
$t<0$ and $|t|$ quite
big, thus $|\omega \eta |\sim |t|^{1-np}\gg1$ if $%
np<1$. This scenario can also make sure adiabatic approximation
satisfied and $|\omega \eta |$ decreases in the contracting phase,
the physical wavelength stretched to super Hubble scale, seed the
energy density perturbation once it reenters the Hubble radius. Thus
the emergence of primordial perturbation can be reduced to the
criteria
\begin{eqnarray}
np &>&1~\text{in an expanding phase } \nonumber\\
np &<&1~\text{in an contracting phase }. \label{np}
\end{eqnarray}

In term of Eq.(\ref{ns}), the possibilities of the scale invariant
spectrum for fixed $z$ are listed in Table.(%
\@Roman1). We set the shaded region as the Case A, which satisfies
$\frac{3n-1}{pn-1}>0$, while the white region as Case B, which
satisfies $\frac{3n-1}{pn-1}<0$. We take $z=1$ as a simple example
for analysis. In order to get scale invariant primordial
perturbation in an expanding phase, $p>1/n$ is required at first.
When $p=1$, then in Case A, the result with normal dispersion
relation is recovered, i.e. scale invariance only for $n\gg 1$. We
can conclude that accelerated expansion is need if $z=1$ and $p=1$,
this is just inflation. In Case B, the scale invariant spectrum can
also be obtained when $n={2\over 3}$, which corresponds to a period
dominated by matter, while $np<1$ means this period must be a
contracting phase, according to (\ref{np}). There is another
interesting example, i.e. the case with $z=3$ and $p=3$. In this
case, $n_s-1\simeq 0$ for any value of $n$. This means no matter how
background evolves, scale invariant spectrum can be generated. We
can check Case B, and find that the solution $n={1\over 3}$ and
$p=3$ lies in curve $pn=1$ which is prohibited, since the points in
this curve corresponds to condition that $\omega \eta $ remains
constant, and thus fails to cross the sound horizon.

In general, if $p=z\neq 3$, it is impossible to have scale invariant
spectrum for arbitrary $n$, as has been studied in previous works.
However, from Table.(%
\@Roman1), we can see that when $p\neq z$, it is possible to have
scale invariant spectrum with any value of $n$, as long as $p$ and
$n$ satisfy certain constraint. For example, for $z=2$, it can be
found that when $p={1+6n\over 3n}$, which occurs for the expansion
or contraction with Case A, or $p={5-6n\over 3n}$, which occurs for
the expansion or contraction with Case B, the spectrum is scale
invariant. This means that if $n={2\over 3}$, i.e. the phase
dominated by matter, the scale invariance will be obtained in the
expansion with $p=2.5$ or the contraction with $p=0.5$. Thus if
$\omega=f(\eta){k^2\over aM}$, we can have scale invariant spectrum
for the expansion with $f(\eta)\sim 1/a^{3\over 2}$ or the
contraction with $f(\eta)\sim \sqrt{a}$. This result might be
interesting for the studies on the theory of cosmological
perturbations in HL cosmology. In general, it can be showed that if
the detailed balance condition is required, the term with $k^{6}$ in
the equation of perturbation mode, like Eq.(\ref{equ}), exactly
cancels, while the term with $k^{4}$ actually dominates,
e.g.\cite{gaoxian}. However, it seems there is a time dependence in
the $k^4$ term. We show that whether the spectrum is actually scale
invariant is determined by this time dependence, and there can
always be room for the generation of scale invariant spectrum.

\linespread{1.36}
\begin{table}[t]
\centering{\large \
\begin{tabular}{|c|c|cc|}
\hline \multicolumn{2}{|c|}{} & \multicolumn{1}{c|}{$p$} & $p$ \\

\multicolumn{2}{|c|}{} & \multicolumn{1}{c|}{($>1/n$, expansion)} &
($<1/n$, contraction) \\ \hline

\multirow{2}{*}{$z=1$} & $n>1/3$ & \colorbox{Gray}{$~~~~~~~~~~~~\frac{2+3n}{3n}~~~~~~~~~~~~$} & $\frac{4-3n}{3n}%
$ \\ \cline{2-2} & $n<1/3$ & $\frac{4-3n}{3n}$ &

\colorbox{Gray}{$~~~~~~~~~~~~\frac{2+3n}{3n}~~~~~~~~~~~~$} \\
\hline
\multirow{2}{*}{$z=2$} & $n>1/3$ & \colorbox{Gray}{$~~~~~~~~~~~~\frac{1+6n}{3n}~~~~~~~~~~~~$} & $\frac{5-6n}{3n}%
$ \\ \cline{2-2} & $n<1/3$ & $\frac{5-6n}{3n}$ &
\colorbox{Gray}{$~~~~~~~~~~~~\frac{1+6n}{3n}~~~~~~~~~~~~$} \\
\hline
\multirow{2}{*}{$z=3$} & $n>1/3$ & \colorbox{Gray}{$~~~~~~~~~~~~~~3~~~~~~~~~~~~~~$} & $\frac{2-3n}{n}$ \\
\cline{2-2} & $n<1/3$ & $\frac{2-3n}{n}$ &
\colorbox{Gray}{$~~~~~~~~~~~~~~3~~~~~~~~~~~~~~$} \\ \hline
\multirow{2}{*}{$z=4$} & $n>1/3$ & \colorbox{Gray}{$~~~~~~~~~~~\frac{-1+12n}{3n}~~~~~~~~~~~$} & $\frac{7-12n}{%
3n}$ \\ \cline{2-2} & $n<1/3$ & $\frac{7-12n}{3n}$ &
\colorbox{Gray}{$~~~~~~~~~~\frac{-1+12n}{3n}~~~~~~~~~~$} \\
\hline
\end{tabular}
} \caption{The possibilities of the scale invariant spectrum for
fixed $z=1,2,3,4$ are listed.   }
\end{table}
The parameter space for scale invariance can also be represented in
Fig.1 and 2. The corresponding curves of `valid parameter' is
plotted in Fig.1 for Case A and in Fig.2 for Case B. The black line
$np=1$ divide the parameter space into 2 regions. The pink region
denotes those for contracting phases, while the uncolored region
denotes those for expanding phases.
\begin{figure}[tbp]
\centering {\small \centering
\includegraphics[scale=1.2]{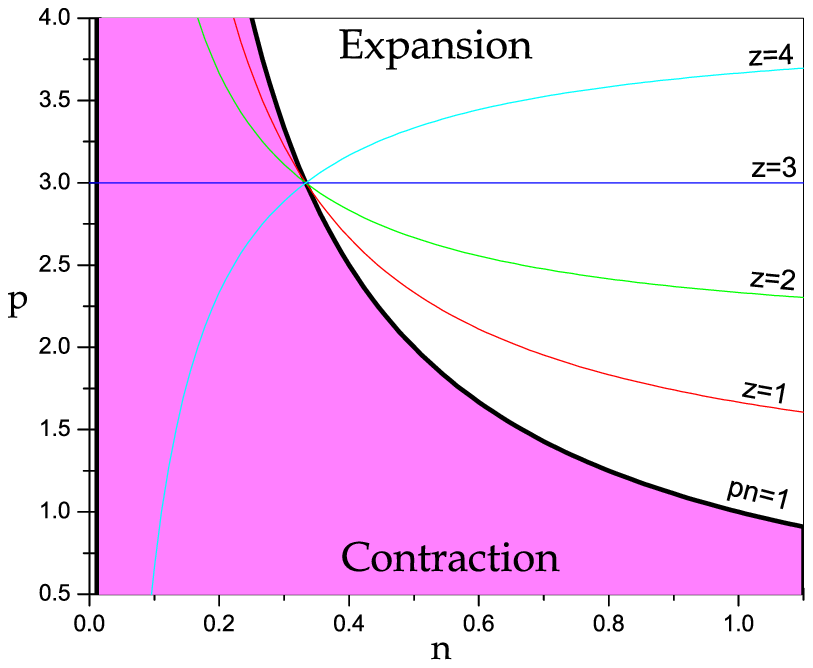} }
\caption{The parameter space for scale invariant spectrum in Case A.
For $z=$1,2,3,4, the constraint for $p$ and $n$ is plotted in the
figure as curves. curves lie in the pink region means corresponding
parameters can only lead to scale invariance in contracting phase,
while any parameters setup represented by point in the uncolored
region can only give scale invariant spectrum in expanding pase. For
an example, we can see that for $z=2$, if $n={2\over 3}$, the
spectrum is scale invariant only when $p=2.5$, which is consistent
with the equation in shaded region in the row with $z=2$ and
$n>{1\over 3}$ in Table.(I).}
\end{figure}

\begin{figure}[tbp]
\centering {\small \centering
\includegraphics[scale=1.2]{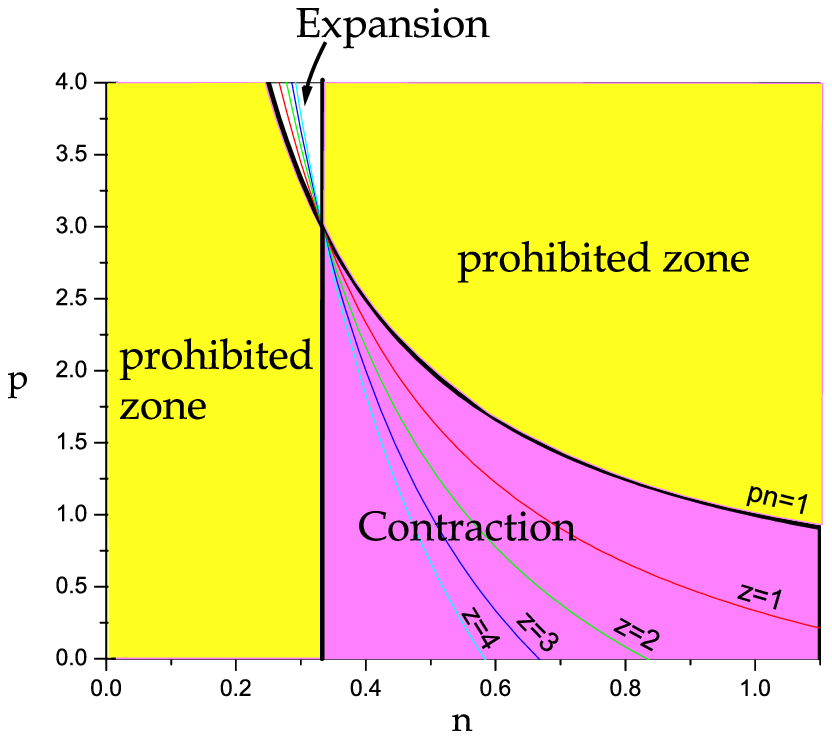} }
\caption{The parameter space for scale invariant spectrum in Case B.
For $z=$1,2,3,4, curves lie in the pink region means corresponding
parameters can only lead to scale invariance in contracting phase,
while any parameter setup represented by point in the uncolored
region can give scale invariant spectrum only in expanding pase.
while the yellow region is prohibited. For an example, we can see
for $z=2$, if $n={2\over 3}$, the spectrum is scale invariant only
when $p=0.5$, which is consistent with the equation in white region
in the row with $z=2$ and $n>{1\over 3}$ in Table.(I).}
\end{figure}

\section{Conclusion}

In this paper, the spectrum of primordial perturbation induced by
a scalar field with modified dispersion relation $\omega\sim
{k^z\over a^{p-1}}$ is studied, in which $z$ is the critical
exponent. We generalize the calculations of \cite{piao09}, in
which $p=z$, to the case with arbitrary $p$ and show the general
result for scale invariant spectrum. We find that when $p\neq z$,
there is always a room of parameters for the generation of scale
invariant spectrum whenever $n$ and $z$ are. It should be pointed
out that here what we concern is only how the primordial
perturbation can emerge and be scale invariant, and whether other
problems of standard cosmology can be solve simultaneously still
needs to be analysed. However, this work can be interesting for
further studies on HL cosmology.

\textbf{Acknowledgments} We thank Y.F. Cai for discussion. This work
is supported in part by NSFC under Grant No:10775180, in part by the
Scientific Research Fund of GUCAS(NO.055101BM03), in part by CAS
under Grant No: KJCX3-SYW-N2.


\end{document}